\documentclass[epj]{svjour}
\usepackage{amsmath}
\usepackage{xspace,graphicx}
\usepackage{cases}
\usepackage{subfigure}

\newcommand{\ie}{{\it i.e.}\xspace}
\newcommand{\eg}{{\it e.g.}\xspace}

\newcommand{\ave}[1]{\left\langle #1 \right\rangle}

\newcommand{\Tref}[1]{Table~\ref{table:#1}}

\newcommand{\tlabel}[1]{\label{table:#1}}

\newcommand{\elabel}[1]{\label{eq:#1}}

\newcommand{\Eref}[1]{Eq.~(\ref{eq:#1})}

\newcommand{\slabel}[1]{\label{sec:#1}}

\newcommand{\Sref}[1]{Section~\ref{sec:#1}}

\newcommand{\flabel}[1]{\label{fig:#1}}

\newcommand{\Fref}[1]{Figure~\ref{fig:#1}}

\newcommand{\OC}{\mathcal{O}}

\newcommand{\imag}{\imath}
\newcommand{\plaind}{\mathrm{d}}

\newcommand{\dint}[1]{\mathchoice{\!\plaind#1\,}{\!\plaind#1\,}{\!\plaind#1\,}{\!\plaind#1\,}}
\newcommand{\half}{\mathchoice{\frac{1}{2}}{(1/2)}{\frac{1}{2}}{(1/2)}}

\renewcommand{\exp}[1]{\mathchoice{e^{#1}}{\operatorname{exp}\left(#1\right)}{\operatorname{exp}\left(#1\right)}{\operatorname{exp}\left(#1\right)}}

\newcommand{\oddSum}{{\text{\tiny odd}}}
\newcommand{\surv}{\sigma}
\newcommand{\escape}{\varepsilon}
\newcommand{\pdf}[1]{\mathcal{P}\left(#1\right)}

\usepackage{dsfont}
\newcommand{\gpset}[1]{\mathds{#1}}

\newcommand{\Zset}{\gpset{Z}}

\newcommand{\latticeSpacing}{a}

\newcommand{\bra}[1]{\left\langle#1\right|}
\newcommand{\braket}[2]{\left\langle#1\middle|#2\right\rangle}
\newcommand{\ket}[1]{\left|#1\right\rangle}

\begin{document}
\title{The average avalanche size in the Manna Model and other
models of self-organised criticality}
\titlerunning{Average avalanche size in SOC}
\author{Gunnar Pruessner}
\institute{Department of Mathematics, Imperial College London, 180
Queen's Gate, London SW7 2AZ, UK}

\abstract{
The average avalanche size can be calculated exactly in a number of 
models of self-organised criticality (SOC). 
While the calculation is straight-forward in one dimension, it
is more involved in higher dimensions and further complicated by the
presence of different boundary conditions and different forms of
external driving. 
Amplitudes of the leading order are determined analytically and
evaluated to obtain analytical references for numerical work.
A subtle link exists between the procedure to
calculate the average avalanche size and the field theory of SOC.
\keywords{Self-organised criticality -- random walks -- escape time --
scaling}
}

\PACS{
{89.75.Da}{Systems obeying scaling laws} \and
{89.75.-k}{Complex systems} \and
{05.65.+b}{Self-organized systems} \and
{05.70.Jk}{Critical point phenomena}
}

\maketitle

\section{Introduction}  
The average avalanche size in models of self-organised criticality (SOC) 
\cite{BakTangWiesenfeld:1987}
is one of the few observables whose scaling frequently is known
exactly even in non-trivial cases. In numerical simulations, it
often plays the
r{\^o}le of a benchmark for convergence
\cite{HuynhPruessnerChew:2011,HuynhPruessner:2012b}. In the one-dimensional version
of many models, even the amplitude of the average avalanche size is
known exactly, as well as the corrections to scaling \cite{Wegner:1972}.
In the following, exact results for the average avalanche size are
collated and extended to higher dimensions and more complicated boundary
conditions.

In a number of conservative SOC models
\cite{BakTangWiesenfeld:1987,Jensen:1998}, particles (height
units or slope units) perform a random walk from the point of being
added to the system until they leave it. In the Manna model
\cite{Manna:1991a} particles
move independently from site to site, so that their trajectory is
\emph{exactly} a random walk. In that model, particles are added
at (normally randomly and independently chosen) sites by the so-called
external drive. If the number of particles at a site exceeds $1$, all
particles are being redistributed independently and randomly to the
nearest neighbours. This process is repeated until no height exceeds the
threshold of $1$ anymore. Sites that do (and the particles located
there) are called active. The totality of redistributions or topplings
constitutes an avalanche. A complete separation of the time scales of
driving and relaxation is achieved by driving only when no site (or
particle) is active. A particle may rest for very long times until being
moved again, but while it is moving, it performs a random walk in a time
that advances only as long as the particle is active (conditional time
scale).

Even in systems where particle movements are not independent, such as
the BTW \cite{BakTangWiesenfeld:1987} and the Oslo Models
\cite{ChristensenETAL:1996}, where
particles are redistributed evenly among nearest neighbours, it has been
noted that particles follow random-walker trajectories, because the entire
ensemble of possible paths is being generated as sites topple.

It has therefore been noted several times that the average avalanche size in the
Manna Model is essentially given by the average escape time of a random
walker, \eg \cite{NakanishiSneppen:1997,NagarPradhan:2003}. While particles in the Manna Model describe
trajectories of a random walker, each of their moves from one site to a
neighbouring site is caused by a
toppling. In fact, in the Abelian version \cite{Dhar:1999c} considered
in the following, each toppling causes two particles ($2d$ particles in
the BTW and Oslo Models on hypercubic lattices\footnote{More generally, in BTW and Oslo Models,
$q$ particles topple, where $q$ is the coordination number of the
lattice in the bulk.}) to move and so the
average number of topplings per particle added, which is exactly the
average avalanche size, is equal to half the average number of moves
each particle makes until its departure from the system. 

The number of
charges a particle causes during its lifetime (\ie the number of times a particle arrives at a 
site until it leaves the system), is exactly equal to the number of
moves it makes; while the initial deposition represents a charge, but
not a move, the final move (off the system) does not cause a charge.

As opposed to higher moments,
the average avalanche size can be calculated because it does not require
any information about the \emph{collective} toppling of
particles.\footnote{In contrast, the present approach does not allow the
calculation of the average avalanche size in the ensemble of avalanches
with non-vanishing size.} It is
merely a matter of stationarity and conservation. The former is
important because only at stationarity the average avalanche size
can be determined as the number of topplings per particle exiting 
by averaging over so many avalanches that the vast
majority of particles added have left the system. Conservation is
important for two reasons. Firstly, particles should not disappear by
interaction, which cannot be accounted for in this simple approach.
Secondly, each and every toppling must count towards an avalanche.

In the following, the average avalanche size is calculated for
hypercubic systems in arbitrary
dimensions (but see \Sref{arbitrary_adjacency}). First, it is calculated for a one-dimensional ``lattice''
with two open boundaries. The result is then generalised to the scaling
in arbitrary dimensions. Doing this exactly and 
on the lattice is a difficult undertaking
\cite{NagarPradhan:2003}, but the aim of the following is to determine
the leading order amplitudes.\footnote{In the following, when
quoting results to leading order the equality sign $\simeq$ will be
used.} 
After taking the continuum limit, they are calculated
for a variety of boundary
conditions. Some special cases are discussed. Finally, the result is
related to some recent field theoretic insights.

\section{One dimension}
In one dimension, the average number of moves can be calculated fairly easily for a
variety of boundary conditions. For brevity, I focus on two open
boundaries (\ie particles leave the system if a toppling site attempts
to deposit a particle on an ``outside'' site). If $x_0$ is the site a particle
is added to by the external drive, then the average number of moves
$m(x_0;L)$ the
particle makes until its departure is given by
\cite{GrimmettStirzaker:1992,Stapleton:2007}
\begin{equation}
m(x;L)=1+\frac{m(x+1;L)+m(x-1;L)}{2}
\elabel{rec_m}
\end{equation}
where the open boundaries are implemented by imposing $m(0;L)=m(L+1;L)=0$,
\ie a Dirichlet boundary condition. Rearranging terms produces a
Poisson equation on the lattice, whose
solution is a simple quadratic, 
\begin{equation}
m(x;L)=x(L+1-x) \ .
\elabel{ave_num_of_moves}
\end{equation}
Summing over the uniform drive (\ie $x_0$ uniformly and randomly taken
from $\{1,\ldots,L\}$) gives 
\begin{equation}
\overline{m}(L)=\frac{1}{L}\sum_{x=1}^L m(x;L)=\frac{(L+1)(L+2)}{6}
\end{equation}
and thus the expectation of the avalanche size (first moment) is exactly
\cite{RuelleSen:1992}
\begin{equation}
\ave{s}=\half \overline{m}(L) = \frac{(L+1)(L+2)}{12} \propto L^2 \ .
\elabel{one_d_exact}
\end{equation}

\subsection{Generalisations}
In higher dimensions, the scaling $\ave{s}\propto L^2$ persists, which
is of course just the usual escape time of a random walker: It explores
the distance $L$ within $L^2$ moves. This argument can be made more
rigorous by noting that if the survival probability after $t$ moves (\ie the probability
of the random walker \emph{not} having reached an open boundary) is
$\surv(t,L)$ in one dimension (for the sake of simplicity, this is the 
probability averaged over the uniform drive), then in higher dimensions
$d$ that probability is simply $\surv(t,L)^d$, because of the
independence of the $d$ directions of possible displacement and the
hypercubic nature of the boundaries.\footnote{If the boundaries are
shaped or structured then the survival in one direction depends on the
coordinate in the other.
Results for that case can be found in \cite{NagarPradhan:2003}.}
In the continuum limit, $t$ is better interpreted as a time,
rather than a number of enforced moves.
The average residence time in $d$ dimension, equal to the
average time to escape $\escape_d$, is thus\footnote{The $\simeq$ sign
applies as $\escape_d(L)/2$ is a continuum approximation of
$\ave{s}_d(L)$, yet $\escape_d(L)$, in the continuum, itself is calculated exactly.}
\begin{multline}
2 \ave{s}_d(L) \simeq \escape_d(L)=
\int_0^{\infty} \dint{t} \surv(t,L)^d \\
= \int_0^{\infty} t \left( -\frac{d}{dt}
\surv(t,L)^d\right)
\elabel{two_s_example}
\end{multline}
where $-\frac{d}{dt} \surv(t,L)^d$ is the probability density of escaping at
time $t$. Its structure reflects the fact that the movement in the $d$
spatial directions is independent; $-\frac{d}{dt} \surv(t,L)$ is the
probability density to escape at time $t$ in one direction, of which
there are $d$ (choices),
and $\surv(t,L)^{d-1}$ is the probability to stay within bounds in the
remaining $d-1$ dimensions.
Here and in the following, the factor $2$ in front of $\ave{s}$ (on the
left of \Eref{two_s_example}) is retained, acting as a reminder of its origin as the
number of particles redistributed in each toppling (the avalanche size
$s$ being measured by the number of topplings). In the BTW and
the Oslo Models, that factor $2$ has to be replaced by the coordination
number of the lattice, $2d$ for a hypercubic one with nearest neighbour
interaction.

Because $\surv(t,L)$ is, by dimensional consistency, bound to be
the dimensionless function $\surv(t/L^2,1)$ it follows that
$\escape_d\propto L^2$, in line with the view that the trajectory of a
random walker is essentially a two-dimensional object
\cite{ItzyksonDrouffe:1997}. Claiming that $t/L^2$ is dimensionless means
being somewhat cavalier about the dimension of the diffusion constant $D$, which in
the present context relates time and number of moves. 
If the walker takes, in each time step, one step in any of the $d$ spatial
directions, the variance of its displacement is $1$. The diffusion
constant, on the other hand, is half the variance of the displacement in each
(independent) spatial direction per time, so that $2Dd=1$ on hypercubic
lattices. There is thus
a slight conceptual difference between the active particles in the Manna
Model on the one hand,
which are forced to move to one of their nearest neighbours, and a
random walker with a certain diffusion constant on the other, which is
subject to random motion in each spatial direction independently.

The survival probability can be calculated quite easily, noting that the
normalised eigenfunctions of $\partial_x^2$ with Dirichlet boundary conditions in
one dimension are $\sqrt{2/L} \sin(x q_n)$ with
$q_n=n \pi/L$, where $n=1,2,\ldots$. With periodic boundary conditions, they are $\exp{x
q_n}$ with $q_n=2 n \pi/L$ and any integer $n$, including $0$ and
negative integers, $n\in\Zset$. As it will turn out below, given the
self-adjoint operator $\partial_x^2$, it is the
presence or absence of the zero mode, \ie the constant eigenfunction
with eigenvalue $0$, which decides over conservation or dissipation and
the structure of the resulting equation for $\ave{s}$.

In one dimension, the probability density function (PDF) of a particle under
Brownian Motion started at $x_0$ with diffusion constant $D$ on an interval with open
boundaries at $0$ and $L$ is thus
\begin{equation}
\pdf{x,t;x_0,L} = 
\frac{2}{L} \sum_{n=1}^\infty \sin(x q_n) \sin(x_0 q_n) e^{-D q_n^2 t}
\elabel{pdf_oneD}
\end{equation}
Since the motion in the different directions is independent, the PDF in
higher dimensions is a product of \Eref{pdf_oneD}. The expected residence time is
given by the integral over time and space, in one dimension
\begin{equation}
\int_0^L\dint{x} \int_0^\infty \!\!\!\! \dint{t}\, \pdf{x,t;x_0,L} = \frac{2}{L}
\sum_{n=1,\oddSum}^\infty \frac{2}{q_n} \sin(x_0 q_n) \frac{1}{D
q_n^2} \ ,
\elabel{residence_as_fct_of_x0}
\end{equation}
where the constraint of $n=1,3,5,\ldots$ in the sum being odd comes from the integral of $\sin(x
q_n)$, which gives $2/q_n$ for odd $n$ and $0$ otherwise. For uniform
drive the escape time is given by the integral
$\int_0^L\dint{x_0} (1/L)$ of
\Eref{residence_as_fct_of_x0}. 
The survival probability, $\surv(t,L)$, on the other hand, is given by
\begin{multline}
\surv(t,L)=\frac{1}{L} \int_0^L\dint{x_0} \int_0^L \dint{x}
\pdf{x,t;x_0,L} \\
= \frac{2}{L^2} \sum_{n=1,\oddSum}^\infty
\frac{4}{q_n^2} e^{-D q_n^2 t}
\end{multline}
and therefore
\begin{multline}
2 \ave{s}_d(L) \simeq \escape_d(L)\\
= \left(\frac{2}{L^2}\right)^d \sum_{n,m,\ldots=1,\oddSum}^\infty
\frac{4}{q_n^2} \frac{4}{q_m^2} \ldots \frac{1}{D(q_n^2+q_m^2+\ldots)}
\elabel{general_ave_s}
\end{multline}
where the sum runs over $d$ different indeces. Dhar's
result for the lattice in $d=2$
\cite[Eq. 21]{Dhar:1990a} is recovered by approximating $\cot(\pi
n/(2L+1))\approx 1/\sin(\pi n/(2L+1)) \approx (2L+1)/(\pi n)$ for large
$L$. By comparison with his results it is clear that in general, on
hypercubic lattices the confluent singularities in the finite size
scaling of $\ave{s}$ are $L^1$, $L^0$ etc.

\subsection{One dimension again}
In the following, a few particular results deriving from
\Eref{general_ave_s} are highlighted. In one
dimension, 
\begin{equation}
2 \ave{s}_1(L) \simeq \frac{2}{L^2} \sum_{n=1,\oddSum}^\infty 
\frac{4 L^2 }{\pi^2 n^2} \frac{2 L^2}{\pi^2 n^2} = 
\frac{L^2}{6}
\end{equation}
using $\sum_{n=1,\oddSum}^\infty 1/n^4 = \pi^4/96$
\cite[Secs. 1.471 and 1.647]{GradshteynRyzhik:2000}, consistent with
\Eref{one_d_exact}. Sums of this type frequently occur in
finite temperature field
theory under the label of Matsubara sums \cite{AtlandSimons:2007}. The latter is associated with
the technique of representing the sum as one over residues,
\begin{equation}
2 \sum_{n=1,\oddSum}^\infty \frac{1}{n^4} =
\frac{1}{2 \pi \imag} \oint_C \frac{1}{z^4} \frac{-\imag \pi}{1+\exp{\imag
\pi z}}
\elabel{orig_Matsubara_sum}
\end{equation}
where the contour $C$ (see \Fref{contour}) encircles each (simple) pole of
$-\imag \pi/(1+\exp{\imag \pi z})$, which are
located at $z=n$ and $z=-n$ ($n$ odd; the parity symmetry is the origin of
the factor $2$ on the left) and have residue $1$. Merging the contours for $z=q_n$ and
$z=-q_n$ and deforming the resulting two contours to enclose the single
pole of order $4$ at $z=0$ produces the desired result, as the 
contour has negative orientation and the  residue is $-\pi^4/48$.

\begin{figure}
\subfigure[Initial arrangement of simple poles to evaluate sum \Eref{orig_Matsubara_sum}.]{\includegraphics[width=0.95\linewidth,bb=180 665 431 719]{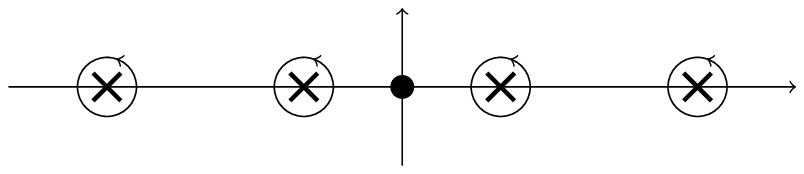}}
\subfigure[Deformation of the initial contour to form two big ones.]{\includegraphics[width=0.95\linewidth,bb=180 665 431 719]{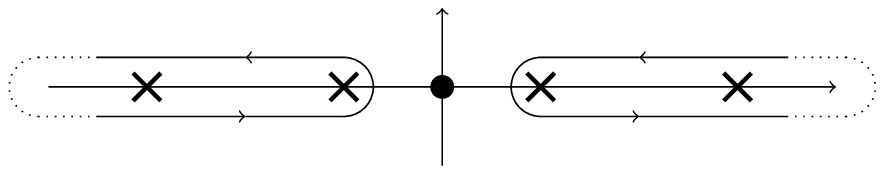}}
\subfigure[Deformation of the contour enclosing a different pole.]{\includegraphics[width=0.95\linewidth,bb=180 665 431 719]{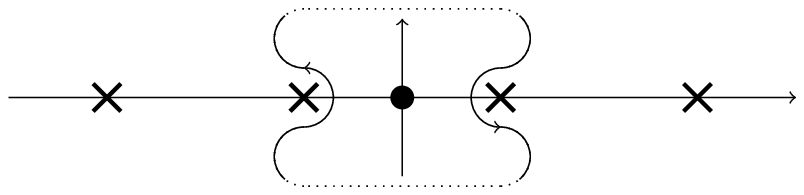}}
\caption{
\flabel{contour} 
Arrangement of the poles in \Eref{orig_Matsubara_sum} in the complex plane and subsequent deformation of the contour.
(a) The sum \Eref{orig_Matsubara_sum} can be performed in a contour integral, by
evaluating $1/z^4$ when calculating the residues at the poles of a suitable
function, indicated by crosses. All circles (with arrows indicating the
direction) together make up the
(initial) contour
$C$ in \Eref{orig_Matsubara_sum}. (b) The contour can be deformed by merging the individual paths. The pole of the
factor $1/z^4$ (filled circle) has to be avoided. The dotted lines indicate the form of the contour for $z\to\pm\infty$. 
(c) Because the integrand drops off sufficiently fast in large arguments, the contours can be joined
up (dotted line) differently, thereby generating a path around the pole of $1/z^4$ with opposite orientation.
}
\end{figure}

It is instructive to attempt to recover \Eref{ave_num_of_moves}, which
is twice the avalanche size for a system driven at site $x=a$. In that
case, the uniform drive, $\int_0^L \dint{x_0} (1/L)$, has to be replaced
by a single source at $a$, \ie $\int_0^L \dint{x_0} \delta(a-x_0)$, so
that the average avalanche size for a system driven at $a$ is
\begin{multline}
2 \ave{s}_{1,a}(L) \simeq \frac{2}{L} \sum_{n=1,\oddSum}^\infty 
\frac{2 L}{\pi n} \sin\left(\frac{n \pi a}{L}\right) \frac{2 L^2}{\pi^2
n^2} \\ 
= \frac{8 L a}{\pi^3} \left\{
\frac{L}{a} \sum_{n=1,\oddSum}^\infty 
\sin\left(\frac{n \pi a}{L}\right) \frac{1}{n^3}
\right\} \ .
\end{multline}
Clearly the terms in the sum contribute significantly less for large
$n$. For small $n$ and large $L$, the $\sin$ may be approximated by its
argument, producing
\begin{equation}
\frac{L}{a} \sum_{n=1,\oddSum}^\infty 
\sin\left(\frac{n \pi a}{L}\right) \frac{1}{n^3}
\approx
\pi \sum_{n=1,\oddSum}^\infty \frac{1}{n^2} 
=
\frac{\pi^3}{8}
\end{equation}
and thus
\begin{equation}
2 \ave{s}_{1,a}(L) \simeq L a
\elabel{better_approx}
\end{equation}
as in \Eref{ave_num_of_moves}. One may be tempted to consider the sum as
Riemann sum with mesh $1/L$, 
\begin{equation}
\frac{L}{a} \sum_{n=1,\oddSum}^\infty
\sin\left(\frac{n \pi a}{L}\right) \frac{1}{n^3}
\approx
\frac{\pi^2 a}{L} \int_{a \pi/L}^\infty  \dint{u}
\sin(u) \frac{1}{u^3}
\approx
\pi
\end{equation}
with dummy variable $u=a n\pi/L$. In the last step, the integrand has
been approximated by $1/u^2$ valid at small $u$. In this approximation,
the avalanche size is
\begin{equation}
2 \ave{s}_{1,a}(L) \approx \frac{8 L a}{\pi^2} \ ,
\end{equation}
a rather poor approximation compared to \Eref{better_approx}.

\section{Two dimensions}
In two dimensions, the same procedures can be followed. For uniform
driving, the key sum to perform is
\begin{multline}
\sum_{n,m=1,\oddSum}^\infty \frac{1}{n^2 m^2 (n^2+m^2)} \\
=
\sum_{m=1,\oddSum}^\infty 
\frac{\pi^2}{8 m^4} - 
\frac{\pi \tanh(\pi m/2)}{4 m^5}\\
= \frac{\pi^6}{768} - \sum_{m=1,\oddSum}^\infty \frac{\pi \tanh(\pi
m/2)}{4 m^5}
\elabel{tanh_summation}
\end{multline}
While the author was unable to determine the last sum (relevant
literature \cite{Apelblat:1996,GradshteynRyzhik:2000,Jolley:1961}), it converges extremely
quickly because of the large power of $m$ in the denominator and because
$\tanh$ very quickly approaches $1$. With the help of Mathematica
\cite{Mathematica:8.0.1.0}, one finds
\begin{equation}
\sum_{m=1,\oddSum}^\infty \frac{\tanh(\pi m/2)}{m^5} = 
0.9216754342259668\ldots
\elabel{numerical_estimate}
\end{equation}
to be compared to $\tanh(\pi/2)=0.91715233\ldots$ and
$\tanh(\pi/2)+\tanh(3 \pi/2)/3^5=0.92126689\ldots$. Using the numerical
estimate \Eref{numerical_estimate}, the average avalanche size in two dimensions with uniform
driving in the bulk and open boundaries is 
\begin{equation}
2 \ave{s}_2(L) \simeq C_2 \frac{64}{D \pi^6} L^2 
\end{equation}
with $C_2=0.5279266525115576573\ldots$ and $D=1/4$, so that
\begin{equation}
\ave{s}_2(L) \simeq 0.070288507477576867\ldots L^2 \ .
\end{equation}

Extensions of the result above to systems with non-unity aspect ratio
$r=L_x/L_y$
are straight forward. The sum to be performed is
\begin{equation}
2\ave{s}_2(L_x,L_y)=\frac{64}{D \pi^6} L_x L_y
\!\!\!\sum_{n,m=1,\oddSum}^\infty 
\frac{1}{n^2 m^2 (r n^2 + r^{-1} m^2)} 
\elabel{ar_orig_expr}
\end{equation}
and thus
\begin{multline}
2\ave{s}_2(L_x,L_y)\\
=\frac{64}{D \pi^6} L_x L_y
\left\{
\frac{\pi^6}{768 r} - \sum_{m=1,\oddSum}^\infty
\frac{\pi \tanh(\pi m r/2)}{4 r^2 m^5}
\right\}
\elabel{ar_rearranged}
\end{multline}
which according to \Eref{ar_orig_expr} is invariant under a change of
$r$ to $r^{-1}$. While this is not at all obvious in
\Eref{ar_rearranged}, the sum is somewhat
reminiscent of that in \cite[Sec 1.471.3]{GradshteynRyzhik:2000}. In the
limit of large $r$, the first term in the curly bracket dominates,
producing $2\ave{s}_2(L_x,L_y)=L_y^2/(12D)$, as the system is
essentially one-dimensional (except for D=1/(2d), due to the additional
degree of freedom). For very small $r$ the sum may be treated as a
Riemann sum.

\subsection{Mixed boundary conditions}
If any of the boundaries is periodically closed or made reflecting,
the dimension (\ie this degree of freedom) effectively disappears from
the problem, \ie the avalanche size is essentially that of a one-dimensional system.
The only trace that remains of the ``closed dimension'' is hidden in the
diffusion constant, which is $D=1/(2d)$, so that
\begin{equation}
\ave{s}_{2,\text{cyl}} = \frac{L^2}{6} + \OC(L)
\elabel{ave_s_cyl}
\end{equation}
for cylindrical boundary conditions on square lattices, $d=2$. In \Eref{ave_s_cyl}
the sub-leading terms are indicated as well, because
\Eref{one_d_exact} remains exact, \ie 
\begin{equation}
\ave{s}_{2,\text{cyl}} =  \frac{(L+1)(L+2)}{6}
\end{equation}
and obviously in higher dimensions
\begin{equation}
\ave{s}_{d,\text{cyl}}=\frac{(L+1)(L+2)d}{12} 
\end{equation}
is the exact expected avalanche size on a hypercubic lattice
if only one direction
remains open, while $d-1$ directions are either periodically closed or
reflecting (or, more generally, produce a spectrum containing $0$).

The technical reason for the simplicity of the results with cylindrical
boundary conditions is the presence of only one sum. The other sums
do not occur because the integration over the entire system as well as
the integration over uniform drive effectively projects the
eigenfunctions of the PDF on a constant, using the scalar product with
constant weight. Under that weight $\nabla^2$ is self-adjoint and
the constant has eigenvalue $0$, provided it is an eigenfunction (which
is decided by the boundary conditions). In that case, the $q_0=0$ mode is
selected in the sum, so that neither any factor $1/q_n$ appears nor a
contribution in $1/(q_n^2+q_m^2+\ldots)$. The mode with eigenvalue $0$
does not decay in time, \ie it is conserved. A boundary condition that
leads to conservation is thus expected to possess such a $0$-mode.

The same type of argument therefore applies in higher dimensions. For
example, when applying periodic boundary conditions to $d-2$
directions in hypercubic
lattices with $d>2$, the average avalanche size is essentially that of a
two-dimensional system, except for $D$ being changed to $D=1/(2d)$.

If individual boundaries have mixed conditions, calculations become
drastically
more complicated. An interesting example is a setup where all boundaries
of a two-dimensional lattice
are reflecting except for a narrow hole of fixed size $h$ from where all particles added
have to escape. Rather counter-intuitively, the scaling of the escape
time in $L$ is not very
different from the scaling on the open lattice, even when the size of
the hole is kept finite and fixed as $L$ is increased. This is surprising,
as the particles need to ``find the narrow exit'' in an increasingly large
system --- given the presence of an additional length scale (the size of
the hole) the scaling of the escape time is no longer determined by
dimensional consistency. On the other hand, one may argue that the
situation is not much different from a one-dimensional lattice, where
the size of the exit remains constant as well. 

On the basis of published results on the narrow escape problem 
\cite{SingerETAL:2006a,SingerETAL:2006b,SingerETAL:2006c}, one
finds
\begin{equation}
2 \ave{s}_{2,\text{narrow}}(L,h) = \frac{2L^2}{\pi D} 
\left\{
\ln\left(\frac{L}{h}\right) + \OC((L/h)^0) 
\right\}
\elabel{narrow_escape}
\end{equation}
where $h$ is the fixed size of the hole adjacent to a corner 
and $D=1/4$ is the diffusion constant. That the size of the hole enters
only very weakly, suggests that the escape time is essentially
determined by the time it takes for the particle to explore the entire
lattice, rather than the size of the exit hole. 
In fact, in dimensions $d\ge2$ a random walker can be thought of as
exploring a convoluted two-dimensional surface with the area covered
(number of
distinct sites visited) increasing essentially linearly in time,\footnote{The walk in one
dimension can be interpreted as a projection from two dimensions.} as if
it was never revisiting a site. Strictly, this holds only in dimensions
strictly greater than two. In two dimensions and less, every site is returned to
infinitely often.

As a final special case in two dimensions, I consider a system driven at
a site with fixed coordinates while the system size is increased. After
the considerations above, it is clear that fixing $d_0\le d$ coordinates
will lead to a scaling $\ave{s}\propto L^{2-d_0}$ for $d_0<2$ in any
dimension $d$. The case $d_0=2$ is special because logarithmic scaling is
expected (whereas a constant average avalanche size occurs for $d_0>2$).
Again, using a Riemann sum introduces uncontrollable errors which are
exacerbated by the sensitivity of the expected logarithm to small
corrections.

The average avalanche size in two dimensions with fixed driving position
is given by
\begin{multline}
2\ave{s}_{2,\text{fixed}} \\
\simeq \frac{1}{D} 
\sum_{n,m=1,\oddSum}^\infty 
\frac{2}{L} \sin(q_n x_0) \frac{2}{q_n}
\frac{2}{L} \sin(q_m y_0) \frac{2}{q_m}
\frac{1}{q_n^2+q_m^2} , 
\end{multline}
where $\sin(n \pi x_0 /L)$ can be approximated by its argument as the terms
in the sum vanish at least like $1/n^3$ in large $n$. For small $n$ the
resulting sum is divergent in the upper limit, which has to be replaced by
the ultraviolet cutoff $L/\latticeSpacing$ with
lattice spacing $\latticeSpacing$,
\begin{equation}
2\ave{s}_{2,\text{fixed}} \approx \frac{16 x_0 y_0}{\pi^2 D} 
\sum_{n,m=1,\oddSum}^{L/\latticeSpacing} 
\frac{1}{n^2+m^2} \ .
\end{equation}
The final result hinges on the last sum. One of the summations can be
performed beyond the upper cutoff without causing a divergence. The
resulting summation involves a term of the form $\tanh(\pi m/2)/m$, which
may be approximated by $1/m$ and thus the sum by
$(\pi/8)\ln(L/(2\latticeSpacing))$, so that
\begin{equation}
2\ave{s}_{2,\text{fixed}} \approx \frac{2 x_0 y_0}{\pi D}
\ln(L/(2\latticeSpacing)) \ .
\end{equation}

The r{\^o}le of the upper cutoff becomes clearer in the case $d_0>2$,
for example fixing the driving position on a three-dimensional lattice.
The reason why the expected escape time remains finite even in the
thermodynamic limit is because within a finite time the random walker, 
attempts to travel beyond the finite distance to one of the open
boundaries,
thus leaving the lattice. Without a finite lattice spacing, the number
of ``hops'' to the open boundary, however, diverges. The difference
between thermodynamic and continuum limit is that absolute distances
correspond to a fixed number of hops in the former, but not in the
latter. From a physical point of view, there is in fact no other difference
between the two.

\section{Higher dimensions}
In higher dimensions the calculation of the relevant sums becomes
increasingly computationally demanding. The expected avalanche size for
homogeneous drive in a $d$ dimensional hypercubic system with open boundaries
generally is according to \Eref{general_ave_s}.
\begin{equation}
2 \ave{s}_d(L) \simeq
\frac{2dL^2}{\pi^2} \left(\frac{8}{\pi^2}\right)^d C_d
\elabel{general_amplitude}
\end{equation}
where $D=1/(2d)$ has been used and
\begin{equation}
C_d = \sum_{n_1,n_2,\ldots,n_d=0}^\infty
\frac{1}{\prod_{i=1}^d (2n_i+1)^2}
\frac{1}{\sum_{i=1}^d (2n_i+1)^2} \ .
\elabel{def_Cd}
\end{equation}
One of the summations can always be carried out, \Eref{tanh_summation}.
Keeping only the two lowest order terms in the resulting sum produces
a recurrence relation for $d>1$,
\begin{multline}
C_d \approx \frac{\pi^2}{8} C_{d-1}
-\frac{\pi}{4(d-1)^{(3/2)}} \tanh\left(\frac{\pi}{2}\sqrt{d-1}\right)\\
-\frac{\pi (d-1)}{36(d+7)^{(3/2)}} \tanh\left(\frac{\pi}{2}\sqrt{d+7}\right)
\elabel{rec_Cd}
\end{multline}
and $C_1=\pi^4/96$ exactly. \Tref{Cd_table} contains the numerical
evaluation of the constants $C_d$ according to \Eref{def_Cd}
together with the approximation \Eref{rec_Cd}. The amplitude in the last
column are well consistent with recent numerical results on the Manna
Model
\cite{HuynhPruessnerChew:2011,HuynhPruessner:2012b}.

\begin{table*}
\begin{center}
\begin{tabular}{l|l|l|l}
$d$ & $C_d$ (numerically) & $C_d$ (approximation \Eref{rec_Cd}) & $(d/\pi^2) (8/\pi^2)^d C_d$ \\
\hline
$1$ & $1.0146780\ldots$ & $1.0146780\ldots$ & $0.0833333\ldots$ \\
$2$ & $0.5279266\ldots$ & $0.5282475\ldots$ & $0.0702885\ldots$ \\
$3$ & $0.3737684\ldots$ & $0.3749565\ldots$ & $0.0605054\ldots$ \\
$4$ & $0.3026980\ldots$ & $0.3055630\ldots$ & $0.0529579\ldots$ \\
$5$ & $0.2651000(1)\ldots$    & $0.2707675\ldots$ & $0.0469927(4)\ldots$ 
\end{tabular}
\caption{
\tlabel{Cd_table}
The constant $C_d$, \Eref{def_Cd}, for dimension $d=1,2,\ldots,5$. 
The second column shows the numerical evaluation of the sum (with extended 
double precision, summing up to $2\cdot1000 +1$ for $d=1,2,3$, up to
$2\cdot500+1$ for $d=4$ and up to $2\cdot200+1$ for $d=5$). 
Unless an error is stated , the digits shown display convergence. The third column is 
the recursive approximation
\Eref{rec_Cd}.
The last column is the amplitude of the leading order $L^2$ of the
average avalanche size, \Eref{general_amplitude}.
}
\end{center}
\end{table*}

\section{Arbitrary Adjacency}
\slabel{arbitrary_adjacency}
\Eref{rec_m} points to a more general procedure to calculate the
expected number of moves to escape from the lattice. If $\ket{m}$ is a
vector whose components $m_i$ are the expected escape times starting
from site $i$ and $A$ is the weighted adjacency matrix (closely related
to Dhar's toppling matrix \cite{Dhar:1990a}, also discussed by Stapleton
\cite{Stapleton:2007}), proportional to the lattice Laplacian, containing $A_{ii}=-1$
across the diagonal and $A_{ij}$ being the probability of $i$
discharging to $j$ (\ie $A_{ij}=1/(2d)$ on hypercubic
lattices),\footnote{Because $A$ does not have to be symmetric, the
procedure described here covers directed models as well.} then
\begin{equation}
-\ket{1}=A\ket{m}
\end{equation}
where $\ket{1}$ is a column of ones. Dissipation at boundary sites is
implemented by $\sum_j A_{ij}<0$, while $\sum_j A_{ij}=0$ at
(conservative) bulk sites. The presence of the non-conservative sites
means that $\ket{1}$ is not an eigenvector, in fact $A\ket{1}$ is a
vector with components that are $0$ for each conservative (bulk) site and negative for
all dissipative (boundary) sites. If $\bra{d}$ is a vector whose components $d_i$ are
the probability that a particle is deposited at site $i$ by the external
drive, with normalisation $\braket{d}{1}=1$, then
\begin{equation}
2\ave{s} = \braket{d}{m}=-\bra{d}A^{-1}\ket{1}
\end{equation}
provided the inverse $A^{-1}$ of $A$ exists. If $A$'s eigenvectors
$\bra{e_i}$ and $\ket{e_i}$ (not necessarily transposed relative to each
other, as $A$ may be directed, \ie not symmetric), with eigenvalues
$\lambda_i$ and $\braket{e_i}{e_j}=\delta_{ij}$, span a subspace
containing $\bra{d}$ and $\ket{1}$ respectively, so that 
\begin{subequations}
\elabel{}
\begin{eqnarray}
\bra{d} &=& \sum_i  u_i \bra{e_i}\\
\ket{1} &=& \sum_i  w_i \ket{e_i}\ ,
\end{eqnarray}
\end{subequations}
then
\begin{equation}
2\ave{s}
= -\sum_i \braket{d}{e_i} \lambda_i^{-1} \braket{e_i}{1} 
=-\sum_i \frac{u_i w_i}{\lambda_i} \ .
\end{equation}
For uniform drive $d_i=1/N$ in a system with $N$ sites and so
$N\bra{d}=\bra{1}$ is a row of ones.  In that case, if $A$ is symmetric 
$u_i=w_i/N$ and
\begin{equation}
2\ave{s} = - \frac{1}{N} \sum_i \frac{\braket{e_i}{1}^2}{\lambda_i} \ .
\end{equation}

\section{Relation to field theory}
There is a subtle but very important link between the calculations
performed above and the field theory of the Manna model
\cite{Pruessner:2012:FT}. Prima facie, it
may look accidental that the calculations for the expectation of the
escape time of a random walker are identical to those for the expected
activity integral. In fact, the bare propagator for the activity at
$\omega=0$ (vanishing frequency, as obtained after Fourier transforming
the time domain) is identical
to that of the time-dependent PDF of the random walker particle.
However, while the former describes the spreading of activity on the
microscopic time scale of the Abelian Manna Model
\cite{Manna:1991a,Dhar:1999c} subject to Poissonian
updates (activated random walkers \cite{DickmanETAL:2000}), the latter describes the movement of a particle on the
conditional time scale, which advances only when the particle is not
stuck on the lattice. Only on that time scale, an actual random walk
is performed and the link exists between the number of moves and the
residence time.

In the light of the field theory, however, it is clear that the particle
movement on the conditional time scale is exactly identical to the
spreading of activity; particles moving are active and vice versa. The
fact that the average avalanche size can be determined by the
considerations presented above means that the bare propagator at
$\omega=0$ is not
renormalised at any order. That does not imply that the bare propagator
is not renormalised at all, as the statement merely applies to
$\omega=0$. In fact, the time dependence of the propagator is very much
expected to be affected by interaction, fluctuations and thus
renormalisation, because active particles do not move freely like a
random walker, but interact with the particles at rest.

The reason why no renormalisation of the propagator at $\omega=0$ takes
place is the same reason that allows the calculation of the average
avalanche size in the first place: Conservation of particles and
stationarity; in the stationary state and because of conservation,
on average exactly one particle leaves the system per particle added.
The number of moves performed by a particle during its residence
determines the average avalanche size.

\section{Acknowledgements}
I would like to thank Abie Cohen, Kyle Johnson, Adam Jones, Xinle Liu
and Yong Won for bringing the narrow escape literature to my attention
and extracting \Eref{narrow_escape}. I also thank Nguyen Huynh for
his suggestions for this manuscript and sharing his insights with me.
Finally, I would like to thank Deepak Dhar for pointing out reference
\cite{NagarPradhan:2003}.

\bibliography{articles,books}

\end{document}